\documentclass[aps,prl,preprint,groupedaddress]{revtex4-1}
\usepackage{bm}
\usepackage{verbatim}   

\usepackage{nopageno}
\usepackage{txfonts}
\usepackage[usenames]{color}

\usepackage{amsfonts,amsbsy,latexsym,amssymb,amscd,amstext}
\usepackage{graphics}
\usepackage{epsfig}
\usepackage{graphicx}
\usepackage{color}      
\usepackage{subfigure}  

\newcommand{\be}{\begin{equation}}
\newcommand{\ee}{\end{equation}}
\newcommand{\ba}{\begin{array}}
\newcommand{\ea}{\end{array}}
\newcommand{\baa}{\begin{array}}
\newcommand{\eaa}{\end{array}}
\newcommand{\bea}{\begin{eqnarray}}
\newcommand{\eea}{\end{eqnarray}}

\newcommand{\G}{G}
\newcommand{\LAMMS}{\Lambda_{\overline{\rm MS}}}

\begin{document}

\title{The string tension from smeared Wilson loops at large N}

\author{  Antonio Gonz\'alez-Arroyo
$^{a,b}$ and Masanori Okawa$^{c}$} 
\affiliation{  $^a$ Instituto de F\'{\i}sica Te\'orica UAM/CSIC \\
  $^b$ Departamento de F\'{\i}sica Te\'orica, C-15 \\
       Universidad Aut\'onoma de Madrid, E-28049--Madrid, Spain \\
       $^c$ Graduate School of Science, Hiroshima University,\\
  Higashi-Hiroshima, Hiroshima 739-8526, Japan
  }

\email{antonio.gonzalez-arroyo@uam.es, okawa@sci.hiroshima-u.ac.jp}

\begin{abstract}
We present the results of a high statistics analysis of smeared Wilson
loops in 4 dimensional  SU(N) Yang-Mills theory for various values of N.
The data is used to analyze the behaviour of smeared Creutz ratios, extracting
from them the value of the string tension and other asymptotic parameters.
A scaling analysis allows us to extrapolate to  the continuum limit for 
N=3,5,6 and 8. The results  are  consistent  with a  $1/N^2$ approach towards 
the large N limit.  The same analysis is done for the TEK model (one-point lattice) for N=841 
and a non-minimal symmetric twist with flux of $k=9$. The results match
perfectly with the extrapolated large N values, confirming the validity
of the reduction idea for this range of parameters. 
\end{abstract}



\preprint{IFT-UAM/CSIC-12-50;\hspace*{3mm} \\
FTUAM-2012-92;\hspace*{3mm}  \\HUPD-1201}

\date{\today}


\maketitle


 
{\vskip 1cm}

There is considerable interest in gauge  theories at large N for
their simplicity, proximity to phenomenologically interesting field
theories  and their presumed connection to string theory.
Lattice gauge theory has proved to be a fundamental tool in deriving 
the non-perturbative properties of Yang-Mills theories at small N. 
In approaching large N, the standard pathway is to study the theory at 
increasing values of N and to extrapolate the results to infinite N. 
This is, no doubt, a costly procedure, with the additional risks involved
in any extrapolation procedure. Nevertheless, results point towards a
somewhat fast approach to the large N limit in many of its
observables~\cite{AGAMO3}-\cite{ATT}. An alternative would be to use the
simplifications involved in the large N theory to find a way to
simulate it directly. An idea going in this direction is that of
{\em reduction} or {\em volume independence}\cite{EK}-\cite{QEK}-\cite{TEK}-\cite{KUY}. 
This allows the possibility of trading the space-time degrees of freedom with those of
the group. The essential ingredient for the idea to work is
invariance under  Z$^4$(N) symmetry, which is broken in the original proposal [3]-[4].
In the twisted Eguchi-Kawai model (TEK) [5], introduced by the present
authors, an invariance  subgroup is preserved at sufficiently weak coupling, 
enabling reduction to work. Recently, it was reported in 
Ref.~\cite{IO}-\cite{TV}-\cite{Az} that  symmetry-breaking takes place 
at intermediate couplings and  N $>$ 100.
To circumvent this problem we proposed a slight variation of the
model~\cite{GAO}.  It exploits the freedom
associated with  an integer parameter entering the formulation, and
representing the chromomagnetic flux  through each two-dimensional plane.
Traditionally this parameter was kept fixed when taking the large N
limit, while we advocated the need to scale it with $\sqrt{N}$ in order to
avoid symmetry-breaking phase transitions. In practice, the
modification involves no additional technical or computational  cost. 
Our initial  tests~\cite{GAO} were free from the problems reported
earlier. To further test  the validity of this idea demanded  performing state of the art
computations of the large N observables and comparing them with those
obtained for the TEK model. Furthermore, even if reduction operates at
the level of the lattice model, our ultimate goal is the
continuum theory, so a scaling analysis is necessary. These were our original
motivations for embarking in the present work. 

Although other observables are possible, we have focused upon the string
tension. This can be obtained as the slope of the linear
quark-antiquark potential. Lately, the best determinations of the
potential and of the string tension have been obtained by
compactifying one dimension, and studying the connected correlation 
function of Polyakov lines\cite{ATT}. In the large N limit this is subleading
with respect to the disconnected term, and it is unclear how to make
the connection. Thus, we stick to the traditional way in which the
string tension is obtained from the expectation value of Wilson
loops $W(T,R)$.    Here one meets a technical but severe difficulty,
since large Wilson loops are very noisy quantities. Furthermore, the 
Wilson loops themselves are affected by ultraviolet divergences so that we will
rather focus on the traditional Creutz ratios:
\begin{equation}
\chi(T,R)= -\log
\frac{W(T+0.5,R+0.5)W(T-0.5,R-0.5)}{W(T+0.5,R-0.5)W(T-0.5,R+0.5)} 
\end{equation}
which are defined for half-integer $R$ and $T$. In the limit $R<<T$
these quantities are lattice approximants to the force $F(R)$ among
quarks separated by a distance $R$. Although, Creutz ratios get rid 
of the constant and perimeter divergences in  Wilson loops, they do
so through  a cancellation, which makes them even more numerically
challenging. To reduce the errors we resort to the well-known
Ape-smearing procedure\cite{Ape} for the ordinary theory. The corresponding 
smearing for the TEK model is given by
\be
U^{smeared}_{\mu}={\rm Proj}_{N}\Bigl[ U_{\mu}+c\sum_{\nu\ne\mu}
(z_{\nu\mu}U_{\nu}U_{\mu}U_{\nu}^{\dagger}
+z_{\mu\nu}U_{\nu}^{\dagger}U_{\mu}U_{\nu})\Bigr] 
\ee
with $z_{\mu\nu}$ the twist tensor. ${\rm Proj}_{N}$ stands for
the operator that projects onto SU(N) matrices.
This process can be iterated several times and  produces a
considerable noise reduction in the data.
One could extract the string tension from the force $F(R)$ obtained 
through   Creutz ratios  for $R<<T$ smeared in the three directions
transverse to $T$. This is, however, very impractical in our case. 
It is much more effective to employ four-dimensional smearing and 
values $R\approx T$. The problem that arises in this approach is that 
not only the error, but also the value of the Wilson loops and Creutz
ratios vary with the  number of smearing steps. This could be an
important source of systematic uncertainties, which might prevent a
precision determination of the string tension from this source. 
To circumvent this problem, our strategy has been to use 
the smeared Creutz ratio values to   {\em extrapolate back} and obtain
un-smeared values. The {\em extrapolated} Creutz ratios do not depend 
on the number of smearing steps, and  the errors are  considerably
smaller than the original un-smeared ratios. Having explained the main observables 
that we will be using, let us summarize in the next paragraph 
the goals and methodology used in this work.

Our main goal is the  determination of the string tension for large N Yang-Mills
theory by means of the study of smeared Creutz ratios on the lattice. 
The large N value will be obtained by extrapolation of data taken 
at N=3,5,6,8 and by direct use of Twisted Eguchi-Kawai model at 
$N=841$ and symmetric twist with flux $k=9$. Indirectly, since the same 
procedures will be used to study the reduced and ordinary model, our results will 
serve to validate the reduction achieved by the TEK model in this
physical range of parameters. Since the goal is the continuum result,
we have simulated the model with Wilson action at several  values of its coupling 
$\beta \equiv 2 N^2 b=2 N^2/\lambda_L$, where $\lambda_L$ is `t Hooft 
coupling on the lattice. The list of parameters and main lattice results are
summarized in Table~I.

\begin{table}
\label{Table I}
\begin{tabular}{||r|r|r|r|r||} \hline \hline
$N$ & $\lambda_L$ \hspace*{3mm} & $u_P$ \hspace*{3mm}  &
$\kappa$\hspace*{8mm} & $\gamma$ \hspace*{6mm}\\ \hline \hline
  3&\ 3.05085&\ 0.58184&\ 0.06737( 75)&\ 0.2110( 22)\\ \hline
  3&\ 3.00000&\ 0.59370&\ 0.04696( 75)&\ 0.2020( 92)\\ \hline
  3&\ 2.95082&\ 0.60414&\ 0.03296( 32)&\ 0.2460( 48)\\ \hline
  3&\ 2.90323&\ 0.61361&\ 0.02471( 33)&\ 0.2309( 64)\\ \hline
  3&\ 2.85714&\ 0.62242&\ 0.01828( 27)&\ 0.2411( 58)\\ \hline
  3&\ 2.81250&\ 0.63064&\ 0.01374( 11)&\ 0.2453(  7)\\ \hline
  3&\ 2.76923&\ 0.63836&\ 0.01055( 12)&\ 0.2399( 14)\\ \hline
  5&\ 2.84625&\ 0.57441&\ 0.04028( 28)&\ 0.2516( 40)\\ \hline
  5&\ 2.78676&\ 0.58892&\ 0.02668( 24)&\ 0.2501( 53)\\ \hline
  5&\ 2.76564&\ 0.59378&\ 0.02244( 20)&\ 0.2662( 15)\\ \hline
  5&\ 2.72242&\ 0.60338&\ 0.01654( 21)&\ 0.2700( 47)\\ \hline
  5&\ 2.65125&\ 0.61836&\ 0.01014( 10)&\ 0.2650( 15)\\ \hline
  6&\ 2.82408&\ 0.56997&\ 0.04126( 35)&\ 0.2580( 75)\\ \hline
  6&\ 2.76923&\ 0.58390&\ 0.02711( 15)&\ 0.2706( 12)\\ \hline
  6&\ 2.74872&\ 0.58883&\ 0.02351( 14)&\ 0.2700( 13)\\ \hline
  6&\ 2.70677&\ 0.59850&\ 0.01729( 12)&\ 0.2747( 29)\\ \hline
  6&\ 2.63736&\ 0.61363&\ 0.01090( 13)&\ 0.2657( 34)\\ \hline
  8&\ 2.80179&\ 0.56548&\ 0.04279( 26)&\ 0.2540( 19)\\ \hline
  8&\ 2.74973&\ 0.57930&\ 0.02840( 17)&\ 0.2672( 13)\\ \hline
  8&\ 2.72869&\ 0.58456&\ 0.02407( 14)&\ 0.2724( 13)\\ \hline
  8&\ 2.68902&\ 0.59405&\ 0.01811( 10)&\ 0.2728( 11)\\ \hline
  8&\ 2.62134&\ 0.60931&\ 0.01112( 12)&\ 0.2726( 33)\\ \hline
841&\ 2.77778&\ 0.55801&\ 0.04234(103)&\ 0.3019(170)\\ \hline
841&\ 2.73973&\ 0.56902&\ 0.03181( 60)&\ 0.2764(118)\\ \hline
841&\ 2.70270&\ 0.57895&\ 0.02474( 56)&\ 0.2623(134)\\ \hline
841&\ 2.66667&\ 0.58805&\ 0.01852( 45)&\ 0.2692( 94)\\ \hline
841&\ 2.63158&\ 0.59651&\ 0.01418( 41)&\ 0.2722( 88)\\ \hline
841&\ 2.59740&\ 0.60442&\ 0.01101( 24)&\ 0.2677( 49)\\ \hline
\end{tabular}
\caption{We list the values of N and lattice couplings $\lambda_L$
studied, together with the plaquette expectation value $u_P$ and best
fit parameters for Eq.~\ref{parameterization}. The N=841 case corresponds to the TEK model.}
\end{table}

The analysis of data and presentation of results follows the
following steps:

{\bf 1.} {\em Measurement of  Wilson loops and 
Creutz ratios.} \\
For N=3, 5, 6, 8 lattice gauge theory, simulations are made on
a $32^4$ lattice with 260 configurations used for each $\lambda_L$.
The number of configurations used in the TEK model for each $\lambda_L$
is 5400, except at $\lambda_L$=2.77778 where it is 2300.
In both the ordinary theory and the TEK model, all configurations are
separated by 100 sweeps, one sweep being defined by one-heat-bath
update followed by five overrelaxation updates.

We determined the Creutz ratios from Wilson loops smeared up to 20
times with $c=0.1$
in the range $R,T\in [3.5,8.5]$. Errors were estimated by jack-knife. 
Smaller values of $R,T$ were also obtained, but dismissed for the
analysis for being more sensitive to lattice artifacts. Larger loops can also
be obtained but are too noisy and/or the number of smearing steps
falls too short for them.

\begin{figure}
\includegraphics{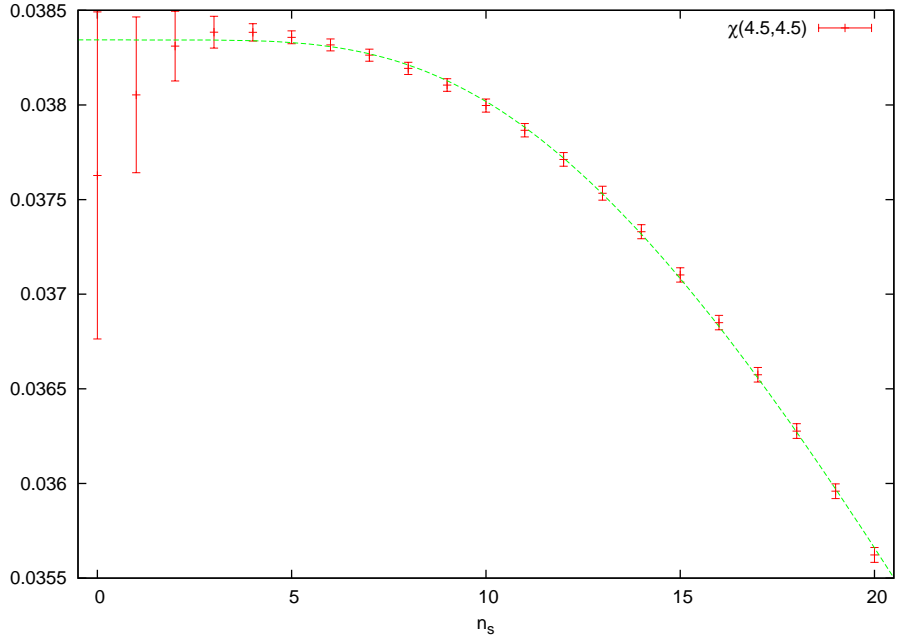}
\caption{$n_s$ dependence of the Creutz ratio $\chi(4.5,4.5)$ for N=841
and $\lambda_L=2.5974$}
\label{fig0}
\end{figure}

{\bf 2.} {\em Extrapolation to the un-smeared Creutz ratio  with error.}\\
The extrapolation procedure depends on the values of $R$ and $T$. For
small values the smeared Creutz ratio are very well fitted to 
a dependence $a(1-\exp\{-b/(n_s+\delta)\})$ where $n_s$ is the number
of smearing steps. This dependence is suggested by perturbation theory.
In Fig.~\ref{fig0} we show, as an example, the $n_s$ dependence of the 
$R=T=4.5$ Creutz ratio for the TEK model and $\lambda_L=2.5974$, together 
with  the corresponding best fit curve. 
For larger values of $R\approx T$ the first smearing steps represent 
a certain transient behaviour, which is then  followed by a plateau, before decaying
consistently with the previous formula. The extrapolated  value is set to the
plateau value. Details apart, it is important to
emphasize that once the protocol to determine the un-smeared Creutz
ratios was defined, it was applied by a program in exactly  the
same way for all values of N and $\lambda_L$, for both the reduced and
ordinary model. Whenever bad fits or ambiguous behaviour was present,
the errors were set to reflect the different options.

{\bf 3.} {\em Analysis of square $R=T$ Creutz ratios}\\
The square Creutz ratios for  large values of $R=T$ are expected to
behave as 
\be 
\label{parameterization}
\chi(R,R)=\kappa+\frac{2\gamma}{R^2}+ \ldots
\ee
A non-zero lattice string tension $\kappa$ is the consequence of
Confinement. The linear term in $1/R^2$ is the predicted behaviour 
both from perturbation theory and 
from a string description of the quark-antiquark flux-tube. The dots
contain corrections from different sources both of continuum and lattice
origin. 

The data of both  models and values of $N$  show a very
clear linear behaviour in $1/R^2$, even at the smallest values of $R$.
A good description of  the data can be obtained with a three parameter fit
based on Eq.~\ref{parameterization}, plus an additional  term of the form
$\eta/R^4$.  The reduced chi square $\sqrt{\chi^2/{\rm ndf}}$ was typically
of order 1 and never exceeded 2. The best fit parameters are listed 
in Table.~I.

{\bf 4.} {\em Scaling analysis}\\
Since our goal is continuum physics we should extrapolate our results
to the continuum limit. Scaling implies that, close enough to the
continuum limit,  results obtained at different 
values $\lambda_L$ should coincide once the lattice spacing $a(\lambda_L)$ is chosen
appropriately. In particular, the length of both sides of a rectangular 
Wilson loop are given in  physical  units  by $t=Ta(\lambda_L)$ and 
$r=Ra(\lambda_L)$. Using this fact and the definition of Creutz ratios 
one concludes: 
\be
\label{scaling_formula}
\chi(T,R)= a^2(\lambda_L) \tilde{F}(t,r) +
\frac{a^4(\lambda_L)}{24}\left(\frac{\partial^2}{\partial
t^2}+\frac{\partial^2}{\partial r^2}\right)  \tilde{F}(t,r) + \ldots
\ee
where the dots contain higher powers of $a(\lambda_L)$. The continuum
function $\tilde{F}(t,r)$ is given by
\be
 \tilde{F}(t,r)= -\frac{\partial^2 \log{\cal W}(t,r)}{\partial r
 \partial t} 
\ee
where ${\cal W}(t,r)$ is the value of the continuum $t\times r$ Wilson
loop. Notice that, although the Wilson loop itself has perimeter and
corner divergences, these disappear when taking the second derivative
with respect to $t$ and $r$. Thus, $\tilde{F}(t,r)$ is a well-defined 
continuum function having the dimensions of energy square. 

In perturbation theory one gets 
\be
\tilde{F}(t,r)=\gamma_P(z)\left(\frac{1}{r^2}+\frac{1}{t^2}\right)
\ee
where $\gamma_P$ is a given function of the aspect ratio $z=r/t$. For the
full non-perturbative theory, one can study the  behaviour of  the
function as $t$ and $r$ goes to infinity. One expects
\be
\label{expansion}
 \tilde{F}(t,r)= \sigma +
 \gamma(z)\left(\frac{1}{r^2}+\frac{1}{t^2}\right) + \ldots 
\ee
where $\sigma$ is the string tension, and the dots represent subleading
terms starting with  $1/(\min(t,r))^4$. The expansion is also exactly
the same as predicted by an effective string theory description of the
Wilson loop expectation value.

This analysis justifies  the parametrization used previously for
square Creutz ratios with $\gamma=\gamma(1)$ and
$\kappa(\lambda_L)=\sigma a^2(\lambda_L)$. In order to compute the
continuum string tension we need to determine $a(\lambda_L)$. For 
very small values of $\lambda_L$ perturbation theory dictates its form:
\be 
\label{apert}
a(\lambda_L)= \frac{1}{\Lambda_L}  \, \exp\{-\frac{1}{2 \beta_0
\lambda_L}\}\  (\beta_0 \lambda_L)^{-\beta_1/(2\beta_0^2)}\equiv
\frac{1}{\Lambda_L} f(\lambda_L)
\ee
However, it is well-known that  scaling seems to work much beyond the region 
where  Eq.~\ref{apert} provides a good approximation. There are several proposals in the
literature, which have been  discussed and tested in many papers, which 
argue that Eq.~\ref{apert} can be extended to the whole scaling 
region using improved couplings  $\lambda_I(\lambda_L)$ in the previous 
formula, instead of $\lambda_L$ itself. All these proposals can be
considered perturbative renormalization prescriptions, and the ratio
of the corresponding scales is obtainable by a perturbative
calculation, i.e. the ratio of lambda parameters. A particular
proposal that has shown good results in previous studies was done by
Parisi~\cite{parisi} and used in the analysis of Ref.~\cite{Edwards}.
When expressed in $\LAMMS$ units it is given by
$a_E=\frac{\LAMMS}{\Lambda_E} f(\lambda_E)$, where
$\lambda_E=(1-u_P)8N^2/(N^2-1)$.
A somewhat different proposal resulted from the analysis of Allton
{\it et~al}.~\cite{ATT}. It is based on a different definition of the effective
coupling $\lambda_A=\lambda_L/u_P$ (and a somewhat modified expression for
$f(\lambda_A)$).

\begin{figure}
\begin{center}
\begingroup
  \makeatletter
  \providecommand\color[2][]{%
    \GenericError{(gnuplot) \space\space\space\@spaces}{%
      Package color not loaded in conjunction with
      terminal option `colourtext'%
    }{See the gnuplot documentation for explanation.%
    }{Either use 'blacktext' in gnuplot or load the package
      color.sty in LaTeX.}%
    \renewcommand\color[2][]{}%
  }%
  \providecommand\includegraphics[2][]{%
    \GenericError{(gnuplot) \space\space\space\@spaces}{%
      Package graphicx or graphics not loaded%
    }{See the gnuplot documentation for explanation.%
    }{The gnuplot epslatex terminal needs graphicx.sty or graphics.sty.}%
    \renewcommand\includegraphics[2][]{}%
  }%
  \providecommand\rotatebox[2]{#2}%
  \@ifundefined{ifGPcolor}{%
    \newif\ifGPcolor
    \GPcolortrue
  }{}%
  \@ifundefined{ifGPblacktext}{%
    \newif\ifGPblacktext
    \GPblacktexttrue
  }{}%
  \let\gplgaddtomacro\g@addto@macro
  \gdef\gplbacktext{}%
  \gdef\gplfronttext{}%
  \makeatother
  \ifGPblacktext
    \def\colorrgb#1{}%
    \def\colorgray#1{}%
  \else
    \ifGPcolor
      \def\colorrgb#1{\color[rgb]{#1}}%
      \def\colorgray#1{\color[gray]{#1}}%
      \expandafter\def\csname LTw\endcsname{\color{white}}%
      \expandafter\def\csname LTb\endcsname{\color{black}}%
      \expandafter\def\csname LTa\endcsname{\color{black}}%
      \expandafter\def\csname LT0\endcsname{\color[rgb]{1,0,0}}%
      \expandafter\def\csname LT1\endcsname{\color[rgb]{0,1,0}}%
      \expandafter\def\csname LT2\endcsname{\color[rgb]{0,0,1}}%
      \expandafter\def\csname LT3\endcsname{\color[rgb]{1,0,1}}%
      \expandafter\def\csname LT4\endcsname{\color[rgb]{0,1,1}}%
      \expandafter\def\csname LT5\endcsname{\color[rgb]{1,1,0}}%
      \expandafter\def\csname LT6\endcsname{\color[rgb]{0,0,0}}%
      \expandafter\def\csname LT7\endcsname{\color[rgb]{1,0.3,0}}%
      \expandafter\def\csname LT8\endcsname{\color[rgb]{0.5,0.5,0.5}}%
    \else
      \def\colorrgb#1{\color{black}}%
      \def\colorgray#1{\color[gray]{#1}}%
      \expandafter\def\csname LTw\endcsname{\color{white}}%
      \expandafter\def\csname LTb\endcsname{\color{black}}%
      \expandafter\def\csname LTa\endcsname{\color{black}}%
      \expandafter\def\csname LT0\endcsname{\color{black}}%
      \expandafter\def\csname LT1\endcsname{\color{black}}%
      \expandafter\def\csname LT2\endcsname{\color{black}}%
      \expandafter\def\csname LT3\endcsname{\color{black}}%
      \expandafter\def\csname LT4\endcsname{\color{black}}%
      \expandafter\def\csname LT5\endcsname{\color{black}}%
      \expandafter\def\csname LT6\endcsname{\color{black}}%
      \expandafter\def\csname LT7\endcsname{\color{black}}%
      \expandafter\def\csname LT8\endcsname{\color{black}}%
    \fi
  \fi
  \setlength{\unitlength}{0.0500bp}%
  \begin{picture}(4818.00,3684.00)%
    \gplgaddtomacro\gplbacktext{%
      \csname LTb\endcsname%
      \put(1078,704){\makebox(0,0)[r]{\strut{} 0.5}}%
      \put(1078,1157){\makebox(0,0)[r]{\strut{} 0.51}}%
      \put(1078,1609){\makebox(0,0)[r]{\strut{} 0.52}}%
      \put(1078,2062){\makebox(0,0)[r]{\strut{} 0.53}}%
      \put(1078,2514){\makebox(0,0)[r]{\strut{} 0.54}}%
      \put(1078,2967){\makebox(0,0)[r]{\strut{} 0.55}}%
      \put(1078,3419){\makebox(0,0)[r]{\strut{} 0.56}}%
      \put(1457,484){\makebox(0,0){\strut{} 0}}%
      \put(1951,484){\makebox(0,0){\strut{} 0.02}}%
      \put(2445,484){\makebox(0,0){\strut{} 0.04}}%
      \put(2939,484){\makebox(0,0){\strut{} 0.06}}%
      \put(3433,484){\makebox(0,0){\strut{} 0.08}}%
      \put(3927,484){\makebox(0,0){\strut{} 0.1}}%
      \put(4421,484){\makebox(0,0){\strut{} 0.12}}%
      \put(176,2061){\rotatebox{-270}{\makebox(0,0){\strut{}$\frac{\Lambda_{\overline{\rm MS}}}{\sqrt{\sigma}}$}}}%
      \put(2815,154){\makebox(0,0){\strut{}$\frac{1}{N^2}$}}%
    }%
    \gplgaddtomacro\gplfronttext{%
      \csname LTb\endcsname%
      \put(3434,1317){\makebox(0,0)[r]{\strut{}SU(N)}}%
      \csname LTb\endcsname%
      \put(3434,1097){\makebox(0,0)[r]{\strut{}TEK N=841}}%
      \csname LTb\endcsname%
      \put(3434,877){\makebox(0,0)[r]{\strut{}$0.515(3)+0.34(1)/N^2$}}%
    }%
    \gplbacktext
    \put(0,0){\includegraphics{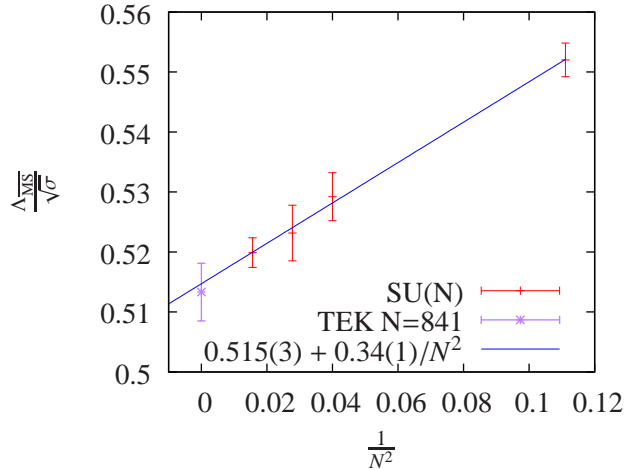}}%
    \gplfronttext
  \end{picture}%
\endgroup
\caption{N dependence of the continuum string tension}
\label{fig1}
\end{center}
\end{figure}

Scaling then implies that the continuum string tension can be
determined in  $\LAMMS$ units as follows:
\be
\frac{\sigma}{\LAMMS^2}=\lim_{a_E\longrightarrow 0}
\frac{\kappa}{a^2_E}
\ee
Our data are consistent with the limit appearing in the right-hand
side of the previous equation being approached linearly in  $a^2_E$.
The extrapolated   values for the ratio $\LAMMS/\sqrt{\sigma}$
 are displayed in Fig.~\ref{fig1} as a function of $1/N^2$. Again 
 a linear fit to the data with parameters $0.515(3)+0.34(1)/N^2$
 is quite satisfactory. The same procedure to obtain the string
 tension in the continuum limit was followed for the TEK model 
 and N=841. The result, also displayed in the figure, is 
 $0.513(6)$. The agreement with the large N extrapolated  value 
 of $\LAMMS/\sqrt{\sigma}$ is very remarkable, 
 and serves as a non-trivial test that reduction is operative 
 for the TEK model in this range. 

 Another remarkable feature of our result is  that the N dependence matches
 perfectly with that obtained in Ref.~\cite{ATT}, which used
 different observables, techniques and  range of `t Hooft couplings. 
 The actual value of the large N ratio given in that reference was 
 0.503(2), which seems inconsistent with our result on statistical grounds.
 However, the estimated systematic errors quoted in Ref.~\cite{ATT} 
 are as large as $0.04$.  We should also mention that a recent analysis,
 largely complementary to ours, has obtained an  estimate of
 the large N string tension which is consistent with our
 result~\cite{lohmayer}.

 In order to give a robust prediction for the large N string
 tension, we should also estimate our systematic errors. The most 
 important source of these errors arises from an overall scale. If 
 we repeat the  procedure replacing the expression of $a_E$ by the
 formula given by Allton {\it et~al}.~\cite{ATT} our estimate of the large N ratio
 $\LAMMS/\sqrt{\sigma}$ becomes $0.525(2)$. This is 
 a 2 percent change in the predicted value, which is 5 times bigger
 than the statistical error.

To give a more precise prediction we should  use a non-perturbative 
renormalization prescription to fix $a(\lambda_L)$. It is possible 
to give a prescription based on Wilson loops and which follows
the same philosophy as the one used to define the Sommer
scale~\cite{sommer}.  Let us consider the
dimensionless function  $\G(r)\equiv r^2 \tilde{F}(r,r)$. A
scale  $\bar{r}$ can be defined as the one satisfying
$\G(\bar{r})=\bar{\G}$. If scaling holds, the choice of $\bar{\G}$ is
irrelevant (provided the
equation has a solution), since it amounts to a change of units.
For our analysis  we took $\bar{\G}=1.65$,  by analogy with Sommer
scale.  However, we checked that taking other choices ($\bar{\G}=2.$
and $\bar{\G}=2.5$) give consistent results up to a change of units.
We recall  that  the idea of considering Creutz ratios with
different aspect ratios $z=R/T$ to define the scale appears in
Ref.~\cite{campostrini}.

One possible way to determine  the scale is by solving for 
$\bar{R}(\lambda_L)$ in the equation
\be
\bar{\G}=\bar{R}^2(\lambda_L)\, \chi(\bar{R}(\lambda_L),\bar{R}(\lambda_L);\lambda_L)
\ee
This gives us $a(\lambda_L)=\bar{r}/\bar{R}(\lambda_L)$. Although, our
data points are defined only for half integer $R$, it is
easy to interpolate and obtain any real $\bar{R}$. Interpolation is a
much more robust procedure than extrapolation, and one can use different
interpolating functions to estimate errors.

The main problem of the previous procedure is that, as explained
previously, the Creutz ratios have intrinsic scaling violations 
given by the second term in Eq.~(\ref{scaling_formula}). Hence, 
a much better  procedure is to make a simultaneous fit to all the square 
Creutz ratio data $\chi(R,R)$ for a particular value of N (and all values
of $\lambda_L$). Combining Eq.~(\ref{scaling_formula}) and the
expansion formula Eq.~(\ref{expansion}), one is led to the following
functional form
\be
\label{fit}
 \frac{\bar{r}^2}{a^2} \chi(R,R;\lambda_L)= \sigma \bar{r}^2 + 2 \gamma \frac{\bar{r}^2}{r^2}+
4\frac{\bar{r}^4}{r^4}(c+d \frac{a^2}{\bar{r}^2 })
\ee
Indeed, this formula describes remarkably well all our data, with 
 chi squares per degree of freedom  of order 1. Notice that the 
coefficient $d$ accounts for the scaling violations arising from the
definition of the Creutz ratios. The remaining terms parametrize the 
continuum function $\tilde{F}(r,r)$ for large values of $r$. This
depends on 3 parameters $\sigma \bar{r}^2$, $\gamma$ and $c$. However,
only two are independent since, by definition, $\bar{r}^2\tilde{F}(\bar{r},\bar{r})=1.65$.  
This fixes $4c=1.65-\sigma \bar{r}^2-2\gamma$. If we substract the term 
proportional to $d$ from the data, all data points should lie in a
universal curve given by the function $\bar{r}^2
\tilde{F}(r,r)=\frac{\bar{r}^2}{r^2} G(r)$. In Fig.~3 we
display the corresponding curve for SU(8) together with the best fit function
extracted from Eq.~(\ref{fit}). Notice how the values obtained from
different couplings fall into a universal curve.  Errors are displayed  
but hard to see at the scale of the graph. Similar curves are obtained for other values
of $N$ and for the TEK model.  

The value of the parameters extracted from the fit are given in
Table~II. Notice that they are very similar for all theories. This 
makes the large N extrapolation very stable. A safe estimate of the 
large N value of $\sigma \bar{r}^2$ is 1.105(10), where the error now 
includes  both statistical and systematic uncertainties. It is clear that 
a good part  of the N dependence and systematic error found before resides in the 
ratio of scales $\bar{r}  \LAMMS$.

\begin{table}
\begin{tabular}{||c|r|r|r|r||} \hline \hline
$N$ & $\sigma \bar{r}^2$ \hspace*{3mm} &  $\gamma$ \hspace*{5mm}
& c \hspace*{6mm} &  d \hspace*{4mm}\\ \hline \hline
  3&\ 1.180( 4)&\ 0.239( 2)&\   -0.0019&\ 0.27( 1)\\ \hline
  5&\ 1.133( 7)&\ 0.263( 3)&\   -0.0023&\ 0.29( 2)\\ \hline
  6&\ 1.120( 4)&\ 0.270( 2)&\   -0.0026&\ 0.30( 1)\\ \hline
  8&\ 1.117( 4)&\ 0.272( 2)&\   -0.0027&\ 0.31( 1)\\ \hline
841&\ 1.130(17)&\ 0.267( 8)&\   -0.0036&\ 0.44( 5)\\ \hline
\end{tabular}
\label{tableII}
\caption{The best fit parameters corresponding to Eq.~\ref{fit}}
\end{table}

In addition to the determination of the string tension, which sets the
long-distance behaviour of Creutz ratios, there is considerable
interest in the parameters that determine the approach to this
long-distance limit. In particular, our results show that the 
large N slope parameter $\gamma$ has a value of $0.272(5)$.  
The slope takes a non-zero value in perturbation theory equal to
$\gamma_P(1)=\frac{(\pi+2)\lambda}{16 \pi^2}(1-1/N^2)$. Using this formula to
define an effective coupling, our data implies $\lambda_{\rm
eff.}\approx 8.4$.   At long distances, however, a new perspective arises
which describes this term as arising from the fluctuation of the chromo-electric flux-tube
stretching among the quark and the anti-quark. In the limit in which 
their separation is large compared to the thickness of
this flux tube, an effective string theory description of the dynamics arises.
The picture predicts~\cite{Luscher:1980fr} that the coefficient of the
$1/r^2$ contribution to the force $F(r)$ is $\gamma(0)=\frac{\pi}{12}$. This
prediction has been verified by lattice data.

\begin{figure}
\begingroup
  \makeatletter
  \providecommand\color[2][]{%
    \GenericError{(gnuplot) \space\space\space\@spaces}{%
      Package color not loaded in conjunction with
      terminal option `colourtext'%
    }{See the gnuplot documentation for explanation.%
    }{Either use 'blacktext' in gnuplot or load the package
      color.sty in LaTeX.}%
    \renewcommand\color[2][]{}%
  }%
  \providecommand\includegraphics[2][]{%
    \GenericError{(gnuplot) \space\space\space\@spaces}{%
      Package graphicx or graphics not loaded%
    }{See the gnuplot documentation for explanation.%
    }{The gnuplot epslatex terminal needs graphicx.sty or graphics.sty.}%
    \renewcommand\includegraphics[2][]{}%
  }%
  \providecommand\rotatebox[2]{#2}%
  \@ifundefined{ifGPcolor}{%
    \newif\ifGPcolor
    \GPcolortrue
  }{}%
  \@ifundefined{ifGPblacktext}{%
    \newif\ifGPblacktext
    \GPblacktexttrue
  }{}%
  \let\gplgaddtomacro\g@addto@macro
  \gdef\gplbacktext{}%
  \gdef\gplfronttext{}%
  \makeatother
  \ifGPblacktext
    \def\colorrgb#1{}%
    \def\colorgray#1{}%
  \else
    \ifGPcolor
      \def\colorrgb#1{\color[rgb]{#1}}%
      \def\colorgray#1{\color[gray]{#1}}%
      \expandafter\def\csname LTw\endcsname{\color{white}}%
      \expandafter\def\csname LTb\endcsname{\color{black}}%
      \expandafter\def\csname LTa\endcsname{\color{black}}%
      \expandafter\def\csname LT0\endcsname{\color[rgb]{1,0,0}}%
      \expandafter\def\csname LT1\endcsname{\color[rgb]{0,1,0}}%
      \expandafter\def\csname LT2\endcsname{\color[rgb]{0,0,1}}%
      \expandafter\def\csname LT3\endcsname{\color[rgb]{1,0,1}}%
      \expandafter\def\csname LT4\endcsname{\color[rgb]{0,1,1}}%
      \expandafter\def\csname LT5\endcsname{\color[rgb]{1,1,0}}%
      \expandafter\def\csname LT6\endcsname{\color[rgb]{0,0,0}}%
      \expandafter\def\csname LT7\endcsname{\color[rgb]{1,0.3,0}}%
      \expandafter\def\csname LT8\endcsname{\color[rgb]{0.5,0.5,0.5}}%
    \else
      \def\colorrgb#1{\color{black}}%
      \def\colorgray#1{\color[gray]{#1}}%
      \expandafter\def\csname LTw\endcsname{\color{white}}%
      \expandafter\def\csname LTb\endcsname{\color{black}}%
      \expandafter\def\csname LTa\endcsname{\color{black}}%
      \expandafter\def\csname LT0\endcsname{\color{black}}%
      \expandafter\def\csname LT1\endcsname{\color{black}}%
      \expandafter\def\csname LT2\endcsname{\color{black}}%
      \expandafter\def\csname LT3\endcsname{\color{black}}%
      \expandafter\def\csname LT4\endcsname{\color{black}}%
      \expandafter\def\csname LT5\endcsname{\color{black}}%
      \expandafter\def\csname LT6\endcsname{\color{black}}%
      \expandafter\def\csname LT7\endcsname{\color{black}}%
      \expandafter\def\csname LT8\endcsname{\color{black}}%
    \fi
  \fi
  \setlength{\unitlength}{0.0500bp}%
  \begin{picture}(8220.00,4818.00)%
    \gplgaddtomacro\gplbacktext{%
      \csname LTb\endcsname%
      \put(946,704){\makebox(0,0)[r]{\strut{} 1}}%
      \put(946,1254){\makebox(0,0)[r]{\strut{} 1.5}}%
      \put(946,1804){\makebox(0,0)[r]{\strut{} 2}}%
      \put(946,2354){\makebox(0,0)[r]{\strut{} 2.5}}%
      \put(946,2903){\makebox(0,0)[r]{\strut{} 3}}%
      \put(946,3453){\makebox(0,0)[r]{\strut{} 3.5}}%
      \put(946,4003){\makebox(0,0)[r]{\strut{} 4}}%
      \put(946,4553){\makebox(0,0)[r]{\strut{} 4.5}}%
      \put(1078,484){\makebox(0,0){\strut{} 0}}%
      \put(2304,484){\makebox(0,0){\strut{} 2}}%
      \put(3531,484){\makebox(0,0){\strut{} 4}}%
      \put(4757,484){\makebox(0,0){\strut{} 6}}%
      \put(5983,484){\makebox(0,0){\strut{} 8}}%
      \put(7210,484){\makebox(0,0){\strut{} 10}}%
      \put(176,2628){\rotatebox{-270}{\makebox(0,0){\strut{}$\bar{r}^2 \tilde{F}(r,r)$  }}}%
      \put(4450,154){\makebox(0,0){\strut{}$2 \frac{\bar{r}^2}{r^2}$}}%
    }%
    \gplgaddtomacro\gplfronttext{%
      \csname LTb\endcsname%
      \put(2926,4380){\makebox(0,0)[r]{\strut{}$\lambda_L=2.80179$}}%
      \csname LTb\endcsname%
      \put(2926,4160){\makebox(0,0)[r]{\strut{}$\lambda_L=2.74973$}}%
      \csname LTb\endcsname%
      \put(2926,3940){\makebox(0,0)[r]{\strut{}$\lambda_L=2.72869$}}%
      \csname LTb\endcsname%
      \put(2926,3720){\makebox(0,0)[r]{\strut{}$\lambda_L=2.68902$}}%
      \csname LTb\endcsname%
      \put(2926,3500){\makebox(0,0)[r]{\strut{}$\lambda_L=2.62134$}}%
      \csname LTb\endcsname%
      \put(2926,3280){\makebox(0,0)[r]{\strut{}Linear part}}%
      \csname LTb\endcsname%
      \put(2926,3060){\makebox(0,0)[r]{\strut{}Quadratic fit}}%
    }%
    \gplbacktext
    \put(0,0){\includegraphics{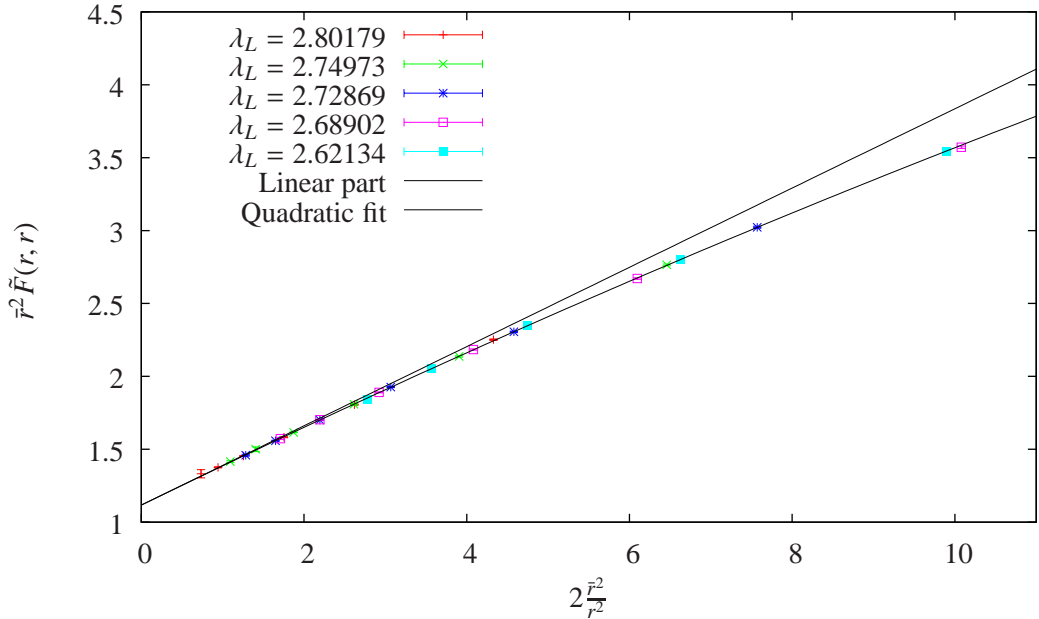}}%
    \gplfronttext
  \end{picture}%
\endgroup
\label{figtres}
\caption{The continuum function $\bar{r}^2 \tilde{F}(r,r)=
\frac{\bar{r}^2}{r^2} G(r)$ is plotted for SU(8)
from our data. The solid lines correspond to the quadratic function
Eq.~\ref{fit}, and to the linear part of this function.} 
\end{figure}

In our present case, it would be possible to study the function 
$\tilde{F}(r,t)$ for $r\ne t$ using information of  non-square smeared
Creutz ratios. In particular the function $\gamma(z)$ provides
interesting information about the properties of the effective string theory.
We can use our data to determine the function $\gamma(z)$ for the
large N theory. The value for $z=1$ coincides with the parameter
$\gamma$ appearing in Table.~II. Since by definition
$\gamma(z)=\gamma(1/z)$, we can parametrize this function in the
vicinity of $z=1$ as $\gamma(z)=\gamma(1)(1+\tau \frac{(z-1)^2}{2z})$.
Our data for $z>0.5$ allow a determination of $\tau$.  For all values
of $N$ and $\lambda$ we get  $\tau=0.31(6)$. 

As mentioned previously, the string picture predicts $\gamma(0)=\pi/12$. 
However, the leading string fluctuation prediction for  $\gamma(1)$ 
is  $\approx 0.16$. Our numerical result for $\gamma(1)$  is 
far from this value and rather close to $\pi/12$. 
The same happens for the $\tau$ coefficient, which is predicted to be
close to 2.  Remarkably, lowest order perturbation theory also has a
prediction for $\tau=2/(\pi+2)\approx 0.39$, which is consistent with
our data. The whole issue of string fluctuations for Wilson loops 
with different aspect ratios is being investigated at
present~\cite{caselle}.

In summary, we have presented a very precise measurement of the
string tension for SU(N) Yang-Mills theory in the large N limit.  It
is remarkable that the N dependence is consistent with that obtained from 
correlation  of Polyakov lines covering a different range of scales and 
distances $r\sqrt{\sigma}$~\cite{ATT}. The large N result is also consistent 
with that obtained from the TEK single-site  model, as predicted by
the reduction idea.





\acknowledgments
A.G-A thanks the GGI Institute,  the organizers
and participants of the 2011 workshop on {\em Large N gauge theories}
for the opportunity to discuss about this topic.  
Financial support from Spanish grants
FPA2009-08785, FPA2009-09017, CSD2007-00042,  HEPHACOS S2009/ESP-1473,
PITN-GA-2009-238353 (ITN STRONGnet) and CPAN CSD2007-00042 is
acknowledged. M. O is supported in part by Grants-in-Aid
for Scientific Research from the Ministry of Education, Culture,
Sports, Science and Technology (No 23540310).

The main calculation has been done on Hitachi SR16000-M1 computer at
High Energy Accelerator Research Organization (KEK) supported by the
Large Scale Simulation Program No.12-01 (FY2011-12).
The calculation has also been done on SR16000-XM1
computer at YITP in Kyoto University and on the INSAM cluster
system at Hiroshima University.


%

\end{document}